\documentclass[aip,reprint,twocolumn]{revtex4-1}

\usepackage{graphicx}
\usepackage{dcolumn}
\usepackage{bm}
\usepackage{color}
\usepackage{amsmath}
\usepackage{amssymb}
\usepackage{graphicx}
\usepackage[unicode=true,pdfusetitle,
 bookmarks=true,bookmarksnumbered=false,bookmarksopen=false,
 breaklinks=false,pdfborder={0 0 0},pdfborderstyle={},backref=false,colorlinks=true]
 {hyperref}
\hypersetup{
 pdfborderstyle={},citecolor=blue}
 
\graphicspath{{./}{./FIGURES/}}

\usepackage[utf8]{inputenc}
\usepackage[T1]{fontenc}
\usepackage{mathptmx}

\begin{document}
\preprint{AIP/123-QED}
 
\title{Bound States of Light Bullets in Passively-Mode-Locked Semiconductor Lasers} 

\author{Fabian Dohmen}
\affiliation{Institute for Theoretical Physics, University of M\"unster, Wilhelm-Klemm-Str. 9, D-48149 Münster, Germany}

\author{Julien Javaloyes}
\affiliation{Departament de F\'{\i}sica, Universitat de les Illes Balears, \& Institute of Applied Computing and Community Code (IAC-3), C/ Valldemossa
	km 7.5, 07122 Mallorca, Spain}	
\author{Svetlana V. Gurevich}
\email{gurevics@uni-muenster.de}
\affiliation{Institute for Theoretical Physics, University of M\"unster, Wilhelm-Klemm-Str. 9, D-48149 M\"unster, Germany}

\affiliation{Center for Nonlinear Science (CeNoS), University of M\"unster, Corrensstrasse 2, D-48149 M\"unster, Germany}
\affiliation{Departament de F\'{\i}sica, Universitat de les Illes Balears, \& Institute of Applied Computing and Community Code (IAC-3), C/ Valldemossa
	km 7.5, 07122 Mallorca, Spain}

\begin{abstract}
In this paper, we analyze the dynamics and formation mechanisms of bound states (BSs) of light bullets in the output of a laser coupled to a distant saturable absorber. 
First we approximate the full three-dimensional set of Haus master equations by a reduced equation governing the dynamics of the transverse profile. This effective theory allows us to perform a detailed multiparameter bifurcation study and to identify the different mechanisms of instability of BSs. In addition, our analysis reveals a non-intuitive dependence of the stability region as a function of the linewidth enhancement factors and the field diffusion. Our results are confirmed by direct numerical simulations of the full system.
\end{abstract}

\maketitle
 
\textbf{Three-dimensional spatio-temporal localized states, so-called light bullets (LBs) were recently predicted theoretically in the output of a laser coupled to a distant saturable absorber. In this paper, we analyze the stability and the range of existence of bound states consisting of LBs. In order to reduce the complexity of the analysis, we first approximate the three-dimensional model by a reduced effective equation governing the dynamics of the transverse profile. This effective theory provides an intuitive picture of the BS formation mechanism. Moreover, it allows us to perform a detailed multiparameter bifurcation study and to identify the different mechanisms of instabilities.} 
 
\section{Introduction}
Localized structures (LSs) appear in a variety of fields ranging from chemical, biological and optical systems to plant ecology and social sciences~\cite{ AA-LNP-08,TML-PRL-94,FS-PRL-96, vanag2007localized,purwins2010dissipative,liehr2013dissipative,meron2015nonlinear,LGRR_JFM_2015,Lloyd201323}. They can interact via exponentially decaying tails and form so-called bound states (BSs) or molecules. BSs can emerge e.g., due to the presence of oscillating tails of the individual interacting LSs, resulting in the formation of structures with stable equilibrium distances and phase differences~\cite{RK-JOSAB-90, Malomed_PRA91, AASC_PRL_97, TMT-PRA-01, moskalenko2003rotational,TVZ_PRE07,GS-LNP-08, SSM_OL18}. If, however, the tails decay monotonically or a non-local repulsive interaction between individual LSs is present, they tend to distribute equidistantly in space or time, leading to periodic pulse trains exhibiting large distances between the consequent LSs~\cite{EMRS_JTB1990,NRV-PD-06,CJM-PRA-16}. However, even if LSs in an individual system exhibit strong repulsion, the formation of BSs can be achieved by arranging several systems in an array with nearest-neighbor coupling~\cite{PVPGY_PRL17}. Recently it was shown that BSs can also be created in systems with a pointwise nonlocality. There, the resulting molecules are composed by LSs which are globally bounded but locally independent~\cite{JMG_PRL17}. 

In optics, LSs of light have been intensively studied theoretically and observed experimentally in both spatial and temporal domains~\cite{FS-PRL-96,BLP-PRL-97,BTB-NAT-02,LCK-NAP-10,HBJ-NAP-14}. In particular, temporal LSs were observed in a semiconductor passively mode-locked laser \cite{MJB-PRL-14}. Passive mode locking (PML) is a well known method for achieving short optical pulses~\cite{haus00rev}. It is achieved by combining two elements inside an optical cavity, a laser amplifier providing gain and a nonlinear loss element, usually a saturable absorber (SA). For proper parameters, this combination leads to the emission of temporal pulses much shorter than the cavity round-trip.
In~\cite{MJB-PRL-14} it was shown that in the so-called long delay limit, where the round-trip time is much longer than the semiconductor gain recovery time, the PML pulses become individually addressable temporal LSs. Interestingly, this temporal localization regime was recently found to be a basis for the generation of the long sought three-dimensional spatio-temporal LSs, so-called light bullets (LBs)~\cite{J-PRL-16}. These pulses of light are simultaneously confined in the transverse and along the propagation directions. Note that in many cases LBs are found to be unstable and collapse in three dimensions~\cite{S-OL-90}, or they were observed in models neglecting the semiconductor dynamics~\cite{VVK-OS-00,BMP-PRL-04}. As such, the LBs reported in \cite{J-PRL-16} are stiff multiple timescale objects since the optical pulse is followed by a material trail that can differ in extension by three orders of magnitudes. This stiffness, stemming from the temporal domain, is aggravated in the presence of the transverse dimensions making a bifurcation analysis of two and three dimensional LBs a challenging problem. The properties of a single LB were investigated in~\cite{GJ-PRA-17}. There, the dynamics of three-dimensional LBs was approximated by the superposition of a slowly evolving transverse profile and a short pulse propagating inside the cavity. This allowed to reduce the dynamics to a partial differential equation governing the transverse profile. A detailed multi-parameter bifurcation study identified several mechanisms of instability where the LBs experience either homogeneous oscillation or symmetry breaking lateral waves radiation. 

In this paper we study the dynamics and formation mechanisms of BSs consisting of LBs. We perform our analysis in two steps: Starting with the three-dimensional generic Haus model, we approximate its solution by the product of a slowly evolving transverse profile and a short pulse propagating inside the cavity. The obtained reduced two-dimensional model, governing the dynamics of the transverse profile is then analyzed employing the continuation and bifurcation package pde2path~\cite{dohnal2014pde2path,pde2path}. Starting with the one-dimensional case that corresponds to the two-dimensional BS of the full system, we show that BSs corresponding to different phase differences can be observed and the stability of these solutions is studied. Our analysis reveals a non-intuitive dependence of the stability region as a function of the linewidth enhancement factors and the field diffusion. In the second stage, our predictions are confirmed by two-dimensional direct numerical simulations of the full system.



\section{Model }

We consider a passively mode-locked laser where a broad area gain chip is coupled to a distant
saturable absorber (SA) with telescopic optics in self-imaging conditions \cite{GBG-PRL-08}. 
We assume that the diffraction in the system, resulting from the propagation within the active sections, is sufficiently small so that 
the paraxial approximation can be used \cite{BLP-PRL-97}. Further, we consider the uniform field limit i.e., the limit of moderate gain ($G$) and saturable absorption
($Q$). In this case, the existence and the dynamical properties of LBs in passively mode-locked VCSELs can be theoretically described \cite{J-PRL-16,GJ-PRA-17} using the generic Haus partial differential equation (PDE) \cite{haus00rev} for the field profile $E\left(r_{\perp},z,\sigma\right)$
over the slow time scale $\sigma$
\begin{eqnarray}
\partial_{\sigma}E & = & \left\{ \sqrt{\kappa}\left[1+\frac{1-i\alpha}{2}G\left(r_{\perp},z,\sigma\right)-\frac{1-i\beta}{2}Q\left(r_{\perp},z,\sigma\right)\right]\right. \nonumber \\
 & - & 1 + \left.\frac{1}{2\gamma^{2}}\partial_{z}^{2}+\left(d+i\right)\Delta_{\perp}\right\} E\left(r_{\perp},z,\sigma\right),\label{eq:VTJ1}
\end{eqnarray}
whereas the carrier dynamics is given by 
\begin{eqnarray}
\partial_{z}G & = & \Gamma G_{0}-G\left(\Gamma+\left|E\right|^{2}\right)+\mathcal{D}_{g}\Delta_{\perp}G,\label{eq:VTJ2}\\
\partial_{z}Q & = & Q_{0}-Q\left(1+s\left|E\right|^{2}\right)+\mathcal{D}_{q}\Delta_{\perp}Q\,. \label{eq:VTJ3}
\end{eqnarray}

Here, $r_{\perp}=\left(x,y\right)$ are transverse space variables, i.e., $\Delta_{\perp}=\partial_{x}^{2}+\partial_{y}^{2}$ is the transverse
Laplacian and the longitudinal variable $\left(z\right)$ can be identified as a fast time variable and represents the evolution of the field within the
round-trip. Further, $\kappa$ is the fraction of the power remaining in the
cavity after each round-trip, $\gamma$ is the bandwidth of the spectral filter
representing, e.g., the resonance of a VCSEL \cite{MJB-JSTQE-15} and $\alpha$ and $\beta$ denote the linewidth
enhancement factors of the gain and absorber sections, respectively. The parameter $d$ represents the small amount of
field diffusion incurred for instance by the dependence of the reflectivity
of the VCSEL distributed Bragg reflectors upon the angle of incidence. In Eqs.~(\ref{eq:VTJ2}-\ref{eq:VTJ3}) $G_{0}$ describes the pumping rate, $\Gamma=\tau_{g}^{-1}$ the gain recovery
rate, $Q_{0}$ is the value of the unsaturated losses, $s$ the ratio
of the saturation energy of the gain and of the SA sections and $\mathcal{D}_{g,q}$
the scaled diffusion coefficients. However, in \cite{J-PRL-16} it was shown 
that carrier diffusion plays almost no role in the LBs dynamics, so that
we set both diffusion coefficients to zero.

We define $G_{th}=\frac{2}{\sqrt{\kappa}}-2+Q_{0}$ as the threshold gain value above which the off solution $\left(E,G,Q\right)=\left(0,G_{0},Q_{0}\right)$
becomes unstable. Then for the gain values below the threshold $G_{th}$ and for proper system parameters Eqs.~(\ref{eq:VTJ1}-\ref{eq:VTJ3}) possess a stable three-dimensional LB solution~\cite{J-PRL-16, GJ-PRA-17}. Note that in general, the non-instantaneous response of the active medium implies a lack of parity along $\left(z\right)$ for the LBs~\cite{J-PRL-16,CJM-PRA-16,GJ-PRA-17,SJG-PRA-18}. 

Using the fact that the LBs are in general composed of variables evolving over widely different timescales, an approximate model governing the shape of the transverse
profile can be derived \cite{N-JQE-74,J-PRL-16,GJ-PRA-17}. In particular, if one considers $E\left(r_{\perp},z,\sigma\right)=A\left(r_{\perp},\sigma\right)p\left(z\right)$,
where $p\left(z\right)$ is a short normalized temporal pulse of length
$\tau_{p}$ that represents the temporal localized state upon which the LB is built
and $A\left(r_{\text{\ensuremath{\perp}}},\sigma\right)$ is a slowly
evolving amplitude, one can separate the temporal evolution into the fast
and slow parts corresponding to the pulse emission and the subsequent
gain recovery. This leads to the following equation governing the dynamics of $A\left(r_{\perp},\sigma\right)$~\cite{J-PRL-16,GJ-PRA-17}:
\begin{eqnarray}
\partial_{t}A & = & (d+i)\Delta A+f\left(\left|A\right|^{2}\right)A.\label{eq:Rosa1}
\end{eqnarray}
Here, the nonlinear function $f=f(|A|^{2})$ reads
\begin{equation}
f\left(I\right) =  \left(1-i\alpha\right)g\left(1+q\right)h\left(I\right)-\left(1-i\beta\right)qh\left(sI\right)-1,\label{eq:Rosa2}
\end{equation}
where $h\left(I\right)=\left(1-e^{-I}\right)/I$, $I=|A|^{2}$.

Similar to the derivation provided in ~\cite{GJ-PRA-17}, the spatial and temporal coordinates in Eqs.~(\ref{eq:Rosa1},\,\ref{eq:Rosa2}) are scaled as $t=\left(1-\sqrt{\kappa}\right)\sigma$
and $\left(x,y\right):=\sqrt{1-\sqrt{\kappa}}\,\left(x,y\right).$ Further, the gain normalized to threshold $g$
and the normalized absorption $q$ are defined as $g=G_{0}/G_{th}$ and $q=Q_{0}/\left(\frac{2}{\sqrt{\kappa}}-2\right)$, respectively.

Note that Eq.~(\ref{eq:Rosa1}) governing the dynamics
of the transverse profile is known in the context of static transverse autosolitons in
bistable interferometers~\cite{RK-OS-88,VFK-JOB-99}. There, the nonlinear
function $h\left(I\right)$ would correspond to a static saturated
nonlinearity, i.e. $f(I)=h\left(I\right)\rightarrow1/(1+I)$. 
By analogy with the static case, and although our nonlinear function is different, we shall call our model for the transverse profile a Rosanov equation.
In this case, for small values of the diffusion coefficient $d$, an interaction law between two weakly overlapping autosolitons was derived in~\cite{VKR_PRE01}. In particular, it was shown that different bound states corresponding to phase differences $\Delta \varphi$  of $0$, $\pi$ and $\pm \pi/2$ exist. Here, the BSs corresponding to the $\Delta \varphi=\{0,\,\pi\}$ are found to be stable if $d$ exceeds certain threshold, whereas for $d=0$ only a BS of $\Delta \varphi=\pi$ corresponding to the minimal distance between both autosolitons is stable. In addition, all BS solutions with $\Delta \varphi=\pm \pi/2$ are found to be unstable. Note that in the case of Eqs.~(\ref{eq:Rosa1},\,\ref{eq:Rosa2}) one can also obtain a similar shape of the nonlinear function by the adiabatic elimination of the carriers, i.e., setting $\partial_{z}G=\text{\ensuremath{\partial}}_{z}Q=0$. However, the reaction time of the gain is known to profoundly affect the stability of spatio-temporal structures \cite{CPM-NJP-06}, that is, its adiabatic elimination approximation along the propagation direction could be incorrect for a semiconductor material. Hence, in what follows we keep the form (\ref{eq:Rosa2}) of the nonlinear function $f(I)$ and perform a bifurcation analysis of BS solutions of Eq.~(\ref{eq:Rosa1}), comparing the results with whose of the Haus PDE~\eqref{eq:VTJ1}-\eqref{eq:VTJ3}.

\section{Results}

\paragraph*{Bifurcation analysis of the Rosanov equation}

We start our analysis with the one-dimensional case. In order to track the localized BS solutions of Eqs.~(\ref{eq:Rosa1}--\ref{eq:Rosa2}) in parameter space, we make use of pde2path~\cite{dohnal2014pde2path,pde2path}, a numerical pseudo-arc-length bifurcation and continuation package for systems of partial differential equations. Taking into account the traslational and the phase-shift symmetries of the system in question, stationary LSs of Eqs.~(\ref{eq:Rosa1}--\ref{eq:Rosa2}) can be found in the form 
\begin{equation}
A(x,t)=u(x-vt)\,e^{-i\omega t}\,,\label{eq:LSSepAnsatz}
\end{equation}
where $u$ is a complex amplitude with the field intensity localized around some
point in space, $\omega$ represents the carrier frequency of
the solution and $v$ is a drift velocity which is in general zero for a BS stationary in time. Substituting Eq.~\eqref{eq:LSSepAnsatz} into Eqs.~(\ref{eq:Rosa1},\,\ref{eq:Rosa2})
we obtain the following equation for unknowns $u$, $\omega$ and $v$
\begin{equation}
\hspace{-0.75cm}(d+i)\partial_x^2 u+v\partial_xu+ i\omega\,u+f\left(\left|u\right|^{2}\right)u=0.\label{eq:RosaOmA}
\end{equation}

\begin{figure}[ht!]
\includegraphics[width=1.0\columnwidth]{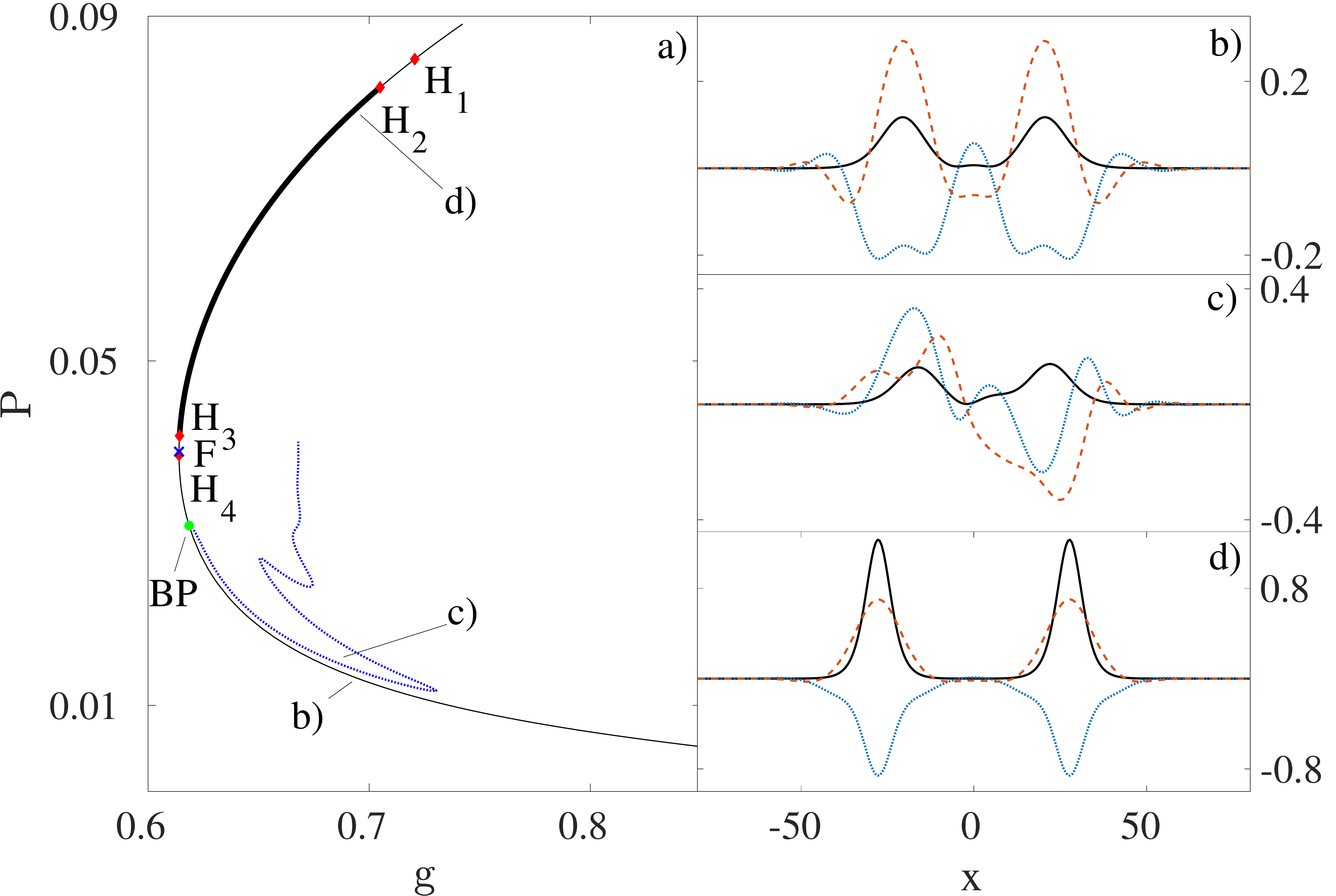}  
\caption{(a) A branch of an one-dimensional BS solution of Eq.\eqref{eq:Rosa1}, calculated for $\Delta\varphi=0$ as a function of the gain parameter $g$. The integrated intensity $P=\int I$ is shown. AH bifurcation points $H_1-H_4$ are marked by red diamonds, whereas the fold ($F$) is marked by a blue cross. A branch of unstable moving BS bifurcates from the branching point (BP, green circle). A BS solution is stable between AH points $H_2$ and $H_3$ (thick line). (b)-(d) Three intensity profiles (black solid lines) corresponding to b) steady unstable BS at $g=0.695$, c) moving unstable BS at $g=0.663$ and d) steady stable BS at $g=0.695$. Blue dotted as well as dashed orange lines correspond to $\Re(u)$ and $\Im(u)$, respectively. Other parameters are ($\gamma,\kappa,\alpha,\beta,q,s,d,L,N_x)=(40,0.8,1.5,0.5,1.27,30,0.03,300,512)$.}
\label{fig:RE1d_PD0} 
\end{figure}

\begin{figure}[ht!]
\includegraphics[width=1.05\columnwidth]{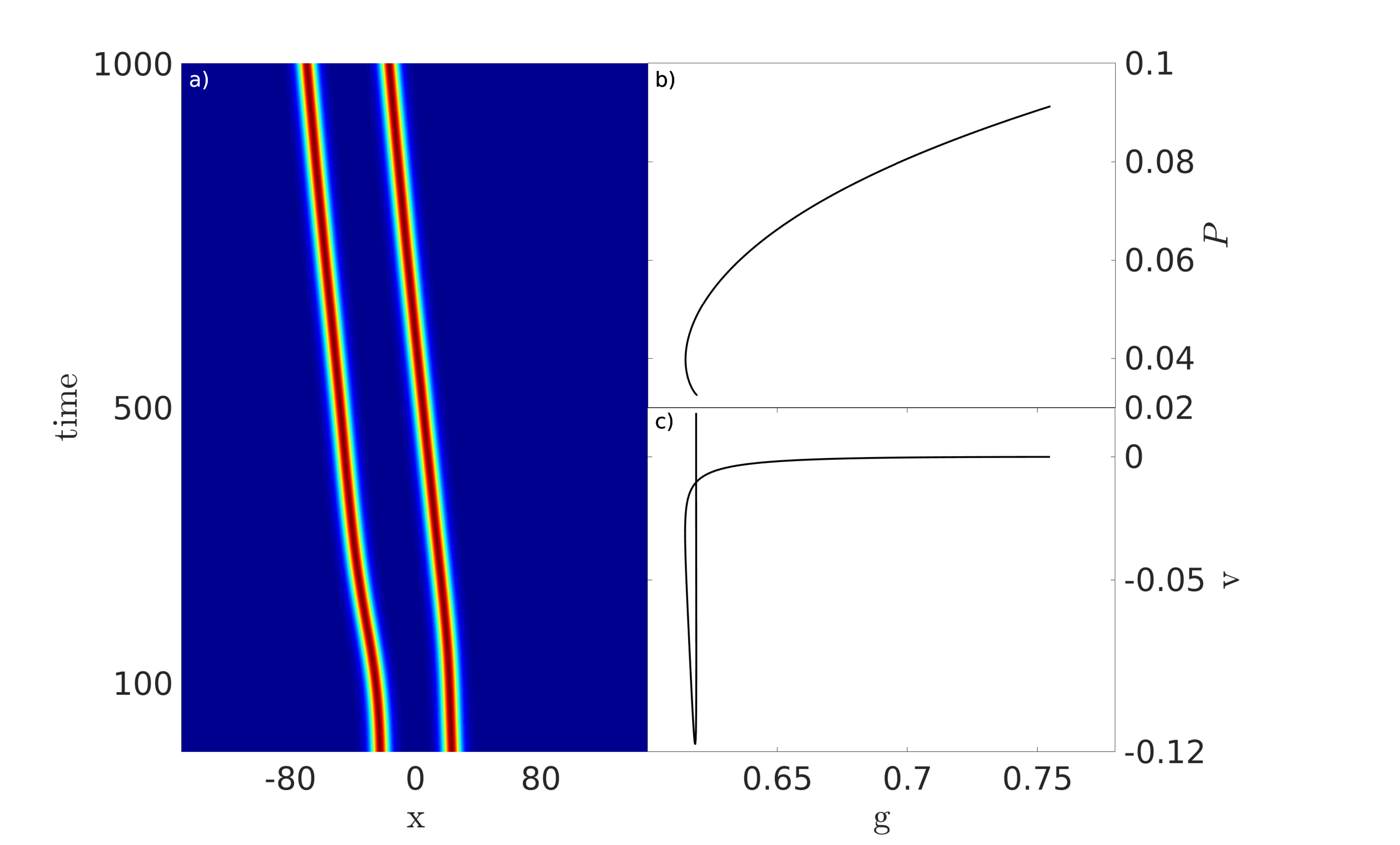}  
\caption{(a) Space-time representation of an one-dimensional moving BS solution of Eq.\eqref{eq:Rosa1} calculated numerically for $\Delta \varphi=\dfrac{\pi}{2}$ at the fixed $g=0.664$. (b) A part of the corresonding solution branch in the $(g,P)$ plane. The whole branch is unstable. (c) The drift velocity of a moving BS as a function of $g$. Other parameters are as in Fig.~\ref{fig:RE1d_PD0}.}
\label{fig:Moving_PDpi2} 
\end{figure}

In the case of Eq.~\eqref{eq:Rosa1} (or \eqref{eq:RosaOmA}), the primary continuation parameter can be, e.g., the gain parameter $g$, the field diffusion $d$ or the linewidth enhancement factors $\alpha, \beta$. However, the spectral parameter $\omega$ and the drift velocity $v$ are two additional free parameters that have to be automatically adapted during the continuation. In order to determine them, we impose additional integral auxiliary conditions
$$
\int\frac{\partial u_{\mathrm{old}}}{\partial r}\,u dx=0\,,
$$
where $u_{\mathrm{old}}$ denotes the solution obtained in the previous continuation step and $r$ represents the spatial coordinate or the phase of the solution. In addition we imposed periodic boundary conditions on the domain of the length $L$ consisting of $N_x$ grid points, i.e. $u(0)=u(L)$. Now one can start a continuation algorithm at, e.g., a numerically given BS solution, continue it in parameter space, and obtain a solution branch. The result of such a continuation for the in-phase BS solution (i.e. the phase difference $\Delta \varphi=0$) is depicted in Fig.~\ref{fig:RE1d_PD0}, where in (a) the integrated intensity $P=\int |u|^2$ as a function of the gain parameter $g$ is shown. One can see that on the high power branch, the BS is unstable w.r.t. Andronov-Hopf (AH) bifurcation $H_1$ for large values of $g$. However, at the next AH point $H_2$ the BS gains the stability and remains stable for a certain interval of $g$ (black thick line). An example of the stable intensity profile (black) is shown in the panel (d) together with $\Re(u)$ (blue dotted) and $\Im(u)$ (orange dashed) of the field. However, the BS looses its stability at further AH point $H_3$ close to the fold $F$. At the lower power branch, the unstable BS solution undergoes the last AH bifurcation at $H_4$ and remains unstable, decreasing its energy with increasing $g$. (cf. Fig.~\ref{fig:RE1d_PD0} (b)). However, at some fixed $g$ value BP (green circle) it experiences another bifurcation. There, a slightly snaking unstable branch of a moving BS solution emerges (see Fig.~\ref{fig:RE1d_PD0}~(c) for the exemplary profile). This branch itself has the multiple AH and fold bifurcation points. Since there are no stable solutions along the branch they are not shown here. 
\begin{figure}[ht!]
\includegraphics[width=1.0\columnwidth]{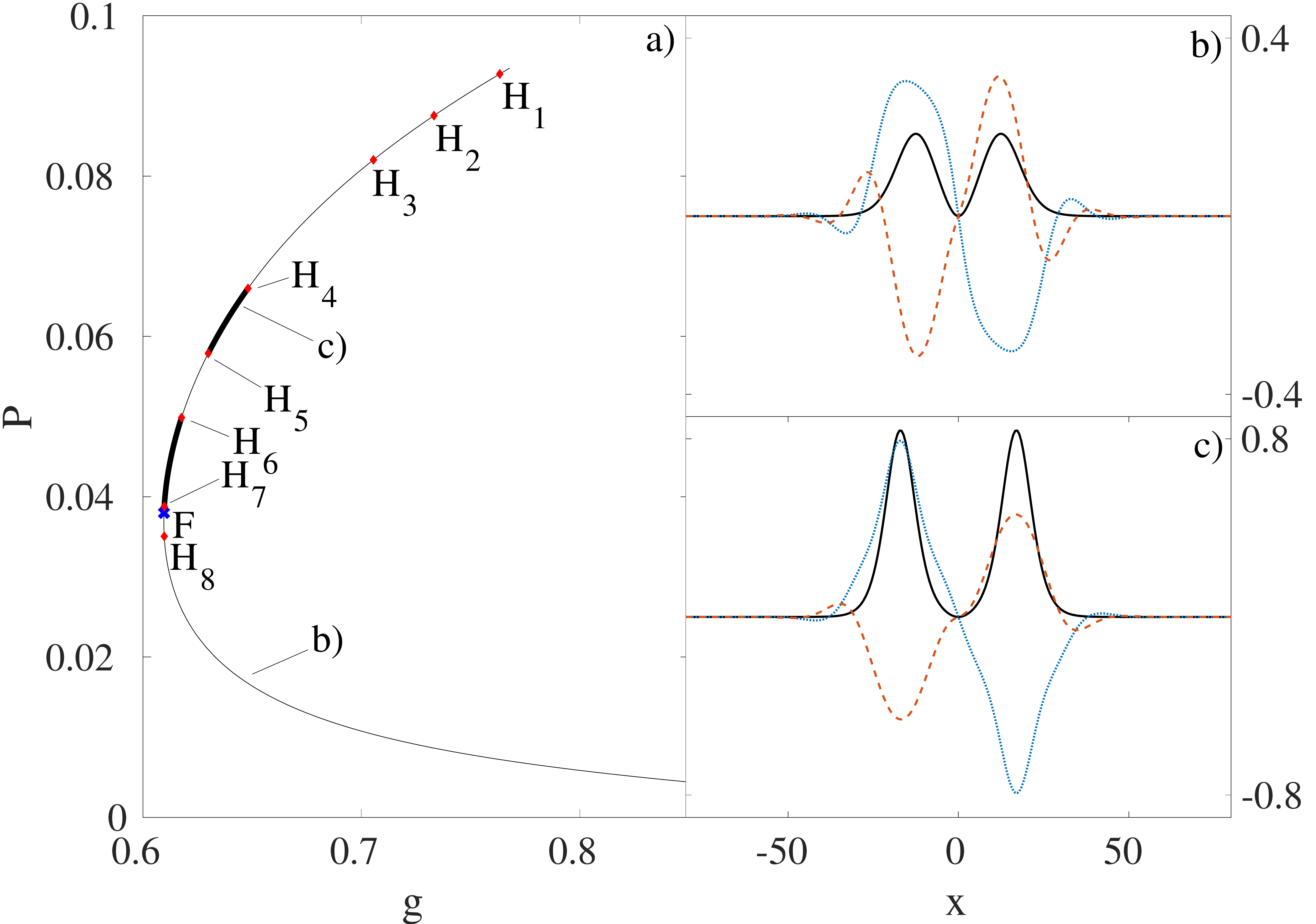}  
\caption{(a) A branch of an one-dimensional BS solution of Eq.\eqref{eq:Rosa1}, calculated for $\Delta\varphi=\pi$ as a function of the gain parameter $g$. The integrated intensity $P=\int I$ is shown. AH bifurcation points $H_1-H_8$ are marked by red diamonds, whereas the fold ($F$) is marked by a blue cross. A BS solution is stable between AH points $H_4$ and $H_5$ as well as between $H_6$ and $H_7$ (thick black line). (b), (c) Two intensity profiles (black solid lines) corresponding to b) steady unstable BS and (c) steady stable BS at $g=0.643$. Blue dotted as well as dashed orange lines correspond to $\Re(u)$ and $\Im(u)$, respectively. Other parameters are as in Fig.~\ref{fig:RE1d_PD0}.}
\label{fig:RE1d_PDpi} 
\end{figure}
In the case of the BS corresponding to $\Delta \varphi=\pi$, the solution branch is shown in Fig.~\ref{fig:RE1d_PDpi}~(a) for the same diffusion value $d=0.03$. Here, the large number of AH bifurcation points arise along the branch. As in the case of $\Delta \varphi=0$, the BS solution is unstable on the high power branch at large $g$ w.r.t. three AH bifurcations $H_1-H_3$ and gains the stability at AH point $H_4$. However, one can see that the region of stability is smaller than in the case of $\Delta \varphi=0$ and is split in two parts, spanned by the pairs of AH points $H_4-H_5$ and $H_6-H_7$. After the fold $F$ and the following AH point $H_8$, the BS solution remains unstable at the low power branch. Note that the branch with unstable moving solutions does not appear for $\Delta \varphi=\pi$. Figure~\ref{fig:RE1d_PDpi} (b), (c) represent two exemplary intensity profiles at $g=0.6436$, corresponding to the unstable and stable BS solutions, respectively. Again, blue dotted as well as dashed orange lines correspond to $\Re(u)$ and $\Im(u)$ of the field. Finally, Fig.~\ref{fig:Moving_PDpi2}~(b) shows the part of the branch of BS solutions calculated for $\Delta \varphi=\dfrac{\pi}{2}$. Although the shape of the branch is similar to the cases of $\Delta \varphi=\{0,\,\pi\}$, this branch does not contain stable time-independent BSs. Instead, the BSs are moving (cf. \ref{fig:Moving_PDpi2}~(a)) with a fixed velocity $v$ depending on the gain value (see Fig.~\ref{fig:Moving_PDpi2}~(c)).

The linear stability of a particular BS solution along the branch can be obtained directly during the continuation so that one has access to the critical eigenfunctions of the system that inform on the particular shape of the wave form. It reveals that the eigenfunctions corresponding to the AH points found for both phase differences $\{0,\,\pi\}$ have different symmetry properties w.r.t. the center of each individual LB as well as w.r.t. the center of the BS. In particular, the AH point $H_2$ ($H_3$) corresponding to $\Delta \varphi=0$ possesses the eigenfunction that is asymmetric (symmetric) w.r.t. the BS center and symmetric (assymetric) w.r.t. each LB. 
\begin{figure}[ht!]
\includegraphics[width=1.05\columnwidth]{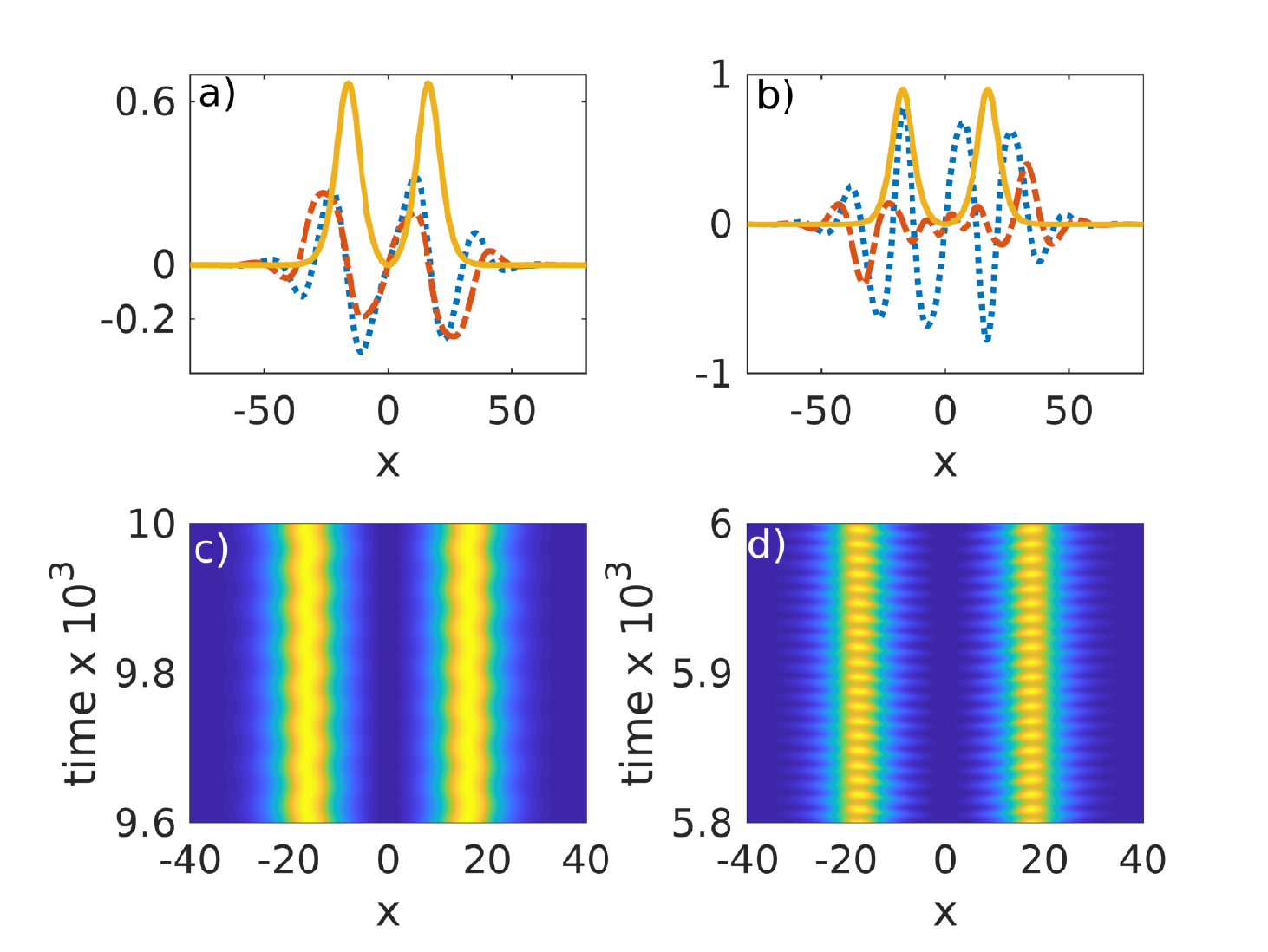}  
\caption{(a,b) Real (blue dotted) and imaginary (orange dashed) parts of critial eigenfunctions associated with the AH points ($H_5, H_4$) together with normalized intensity profile (yellow solid) caclulated for $\Delta \varphi=\pi$ at $g=0.631$ and $g=0.656$, respectively. (c,d) A space-time representation of the intensity field evolution obtained by direct numerical simulations of Eqs.~(\ref{eq:Rosa1}-\ref{eq:Rosa2}) for two different values of $g$ corresponding to (a,b). Other parameters are as in Fig.~\ref{fig:RE1d_PD0}.}
\label{fig:REIMRos} 
\end{figure}
An example of the shapes of critical eigenfunctions corresponding to AH points $H_5, H_4$ for $\Delta \varphi=\pi$ is depicted in Fig.~\ref{fig:REIMRos} (a)-(b), whereas panels (c)-(d) show a space-time representation of the intensity field evolution obtained by direct numerical simulations of Eqs.~(\ref{eq:Rosa1}-\ref{eq:Rosa2}) for two different values of $g$ close to the corresponding AH bifurcation points $H_5, H_4$. As it can be seen in Fig.~\ref{fig:REIMRos}~(a), both real and imaginary parts of the critical eigenfunction are asymmetric towards the center of the BSe  w.r.t. the individual LB centers. Hence, an asymmetrical breathing in space would be expected which is verified by direct numerical simulations represented in the panel (c). In Fig.~\ref{fig:REIMRos}~(b), on the contrary, the real part of the eigenfunction is asymmetric towards the center and is nearly symmetrical in the real part and asymmetrical in the imaginary part w.r.t. the peak maxima. Based on the value of the real and imaginary contributions one would assume a breathing oscillation with an asymmetry in the pulse envelope, cf. Fig.~\ref{fig:REIMRos}~(d). There, a clear pulsation in time is observable as well as a slight elongation of the pulse towards the center. 
\begin{figure}[ht!]
\includegraphics[width=1.1\columnwidth]{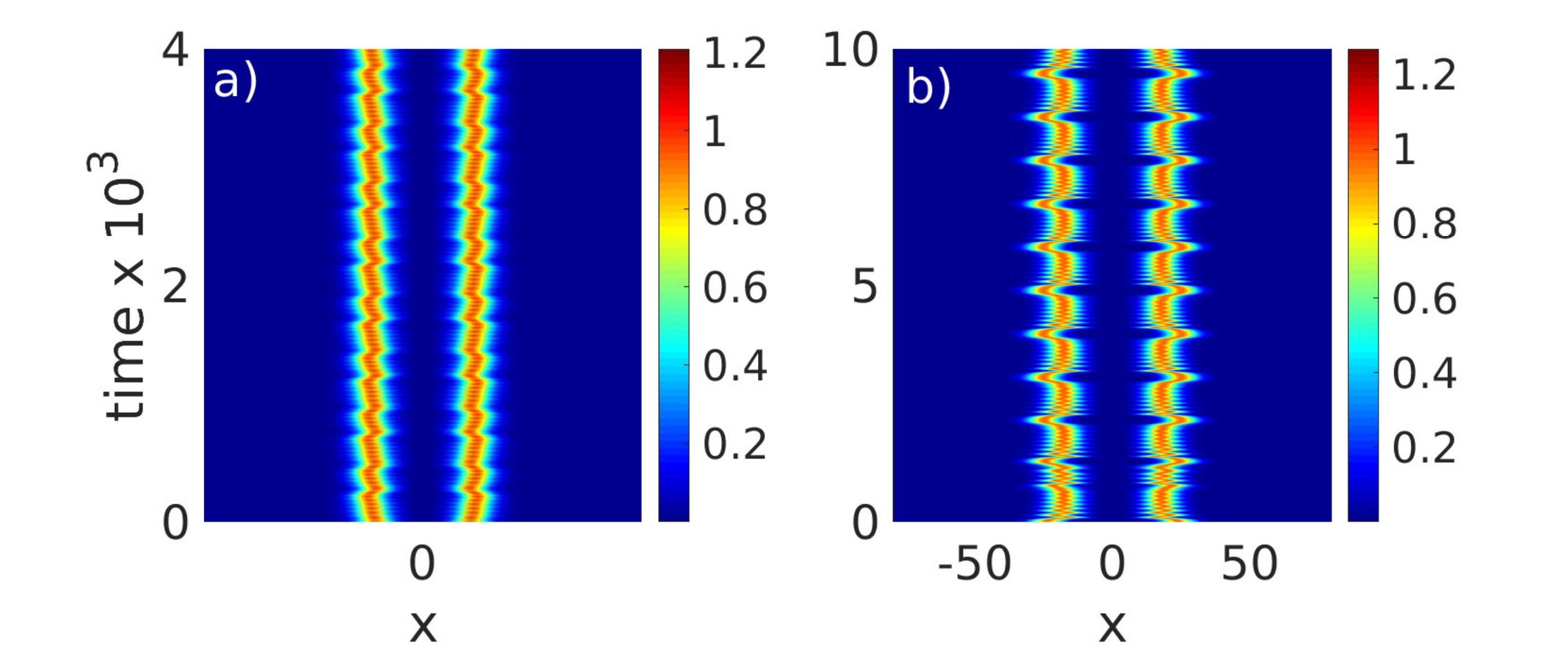}  
\caption{Complex spatio-temporal oscillatory dynamics of a BS solution observed in the high intensity branch at (a) $g= 0.668$ and (b) $g=0.664$ between AH bifurcation points $H_4$ and $H_3$ for $\Delta \varphi=\pi$. Other parameters are as in Fig.~\ref{fig:RE1d_PD0}.}
\label{fig:ComplDyn} 
\end{figure}
The oscillatory dynamics of the BS can however become more complex for increasing values of $g$. Figure~\ref{fig:ComplDyn} shows two examples of the temporal evolution of the BS observed between $H_4$ and $H_3$ bifurcation points. Here, several frequencies contribute to the dynamics leading to a complex spatio-temporal behavior. The numerical continuation analysis reveals the hysteresis behavior for increasing (decreasing) values of $g$ indicating the subcritical character of the AH bifurcation $H_3$.  

Besides the gain $g$, the linewidth enhancement factors of the gain and the absorber sections as well as the field diffusion are important control parameters that determine the stability range of the BS. In order to study the influence of $\alpha$, $\beta$ and $d$ on the stability range of the single BS solution we perform a two-parameter continuation of the fold and AH bifurcation points for the both phase differences $\Delta \varphi=0,\,\pi$.
%
\begin{figure}[ht!]
\includegraphics[width=1.0\columnwidth]{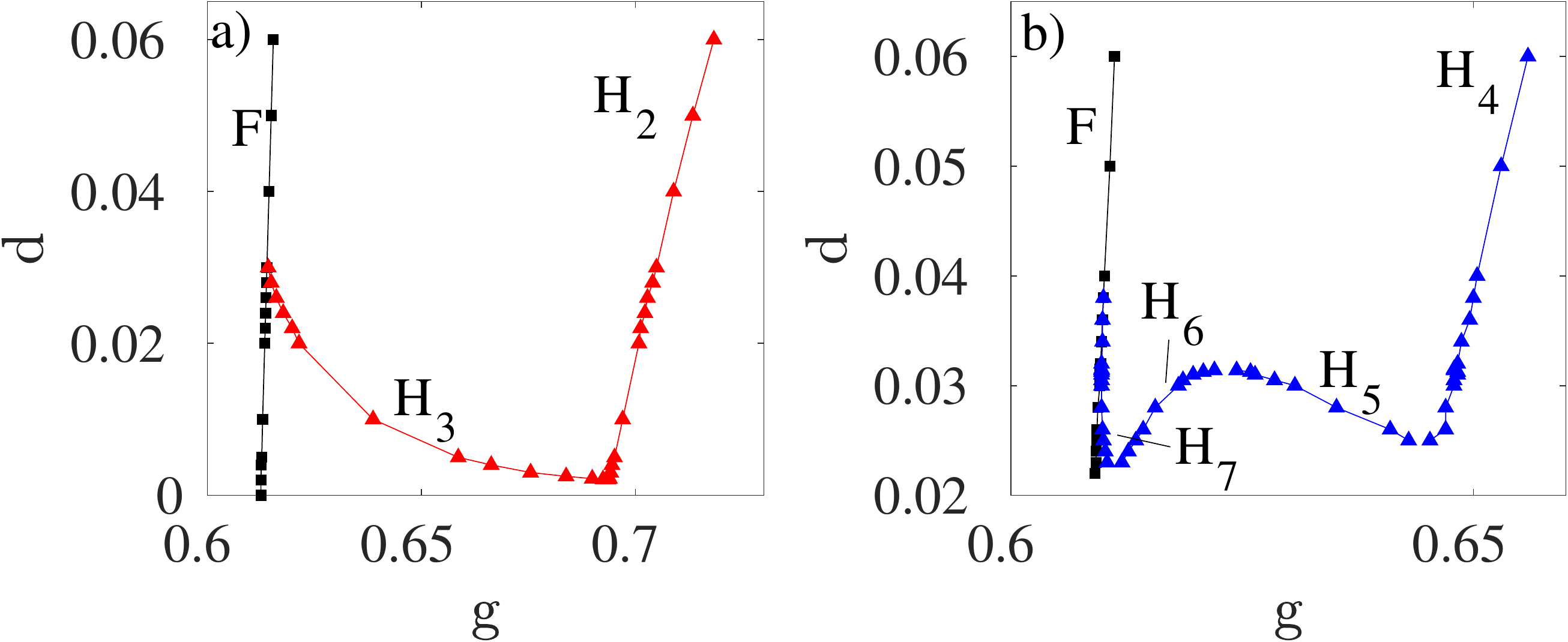}  
\caption{Evolution of the fold (black squares) and AH points $H_2-H_7$ in the $(g, d)$ plane for the phase differences (a) $\Delta \varphi=0$ (red triangles) and  (b) $\Delta \varphi=\pi$ (blue triangles) (cf. Figs.~\ref{fig:RE1d_PD0}, \ref{fig:RE1d_PDpi}). No stable BS solutions exist below the threshold $d_{th}$ for the field diffusion with the minimal value of (a) $d_{th}\simeq 0.002$  and (b) $d_{th}\simeq 0.022$, respectively.
Other parameters are as in Fig.~\ref{fig:RE1d_PD0}.}
\label{fig:Ros1d_diff} 
\end{figure}
Figure~\ref{fig:Ros1d_diff}~(a) shows the stability diagram in $(g,\,d)$ plane for $\Delta \varphi=0$. Here, black squares indicate the fold $F$ threshold, whereas red triangles stand for the boundary of the AH bifurcations $H_2$ and $H_3$. The BS solution is stable in the region bounded by the $F$, $H_2$ and $H_3$ lines. One can see that the AH point $H_3$ moves in the direction of the fold $F$ for increasing values of $d$, so that above a certain threshold, the stability region is bounded by $F$ and $H_2$ lines only. Notice that the stability region is limited from below in $d$, that is, there is no stable BS solutions exist below a certain field diffusion threshold $d_{th}$ depending on the gain value $g$. The stability diagram for the phase difference $\Delta \varphi=\pi$ is depicted in Fig.~\ref{fig:Ros1d_diff}~(b). Here, the stability region for the BS solution is limited by the line given by the fold position $F$ (black squares) and the boundaries of the AH bifurcations $H_4$, $H_5$ and $H_6$, $H_7$ (blue triangles). For increasing values of $d$, the AH point $H_7$ moves in the direction of the fold $F$, whereas $H_6$ and $H_5$ move toward each other until they collide at some critical value of $d$. That is, for large values of the field diffusion the BS is stable between $F$ and $H_4$ lines and only one stability region exists. Note that as in the case of $\Delta \varphi=0$, there exist a threshold value of the field diffusion that bounds the stability region from below. This threshold value also depends on the gain value and is in general higher then in $\Delta \varphi=0$ case. Note that this result is different from the case of static saturated nonlinearity discussed in~\cite{VKR_PRE01}, where the antiphase BS can be stable for vanishing values of the field diffusion.
\begin{figure}[ht!]
\includegraphics[width=1.0\columnwidth]{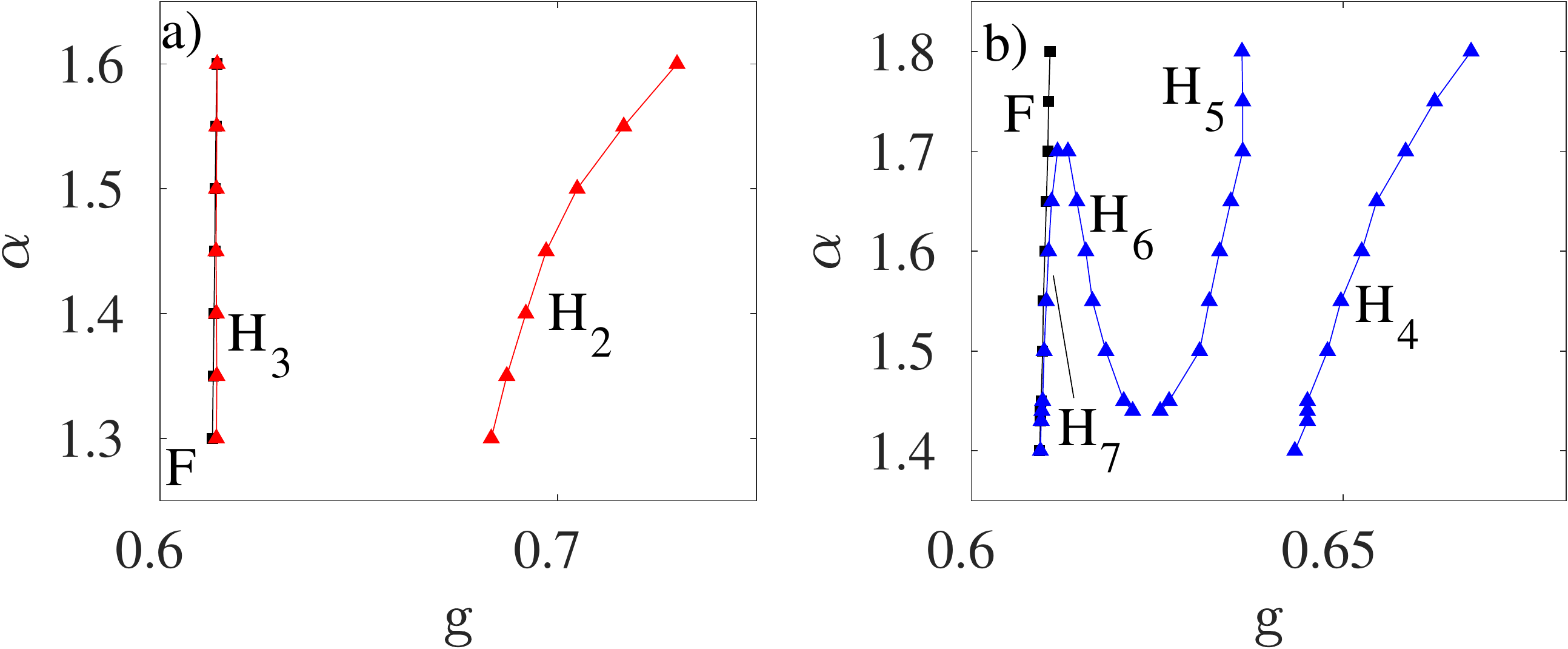}  
\caption{One-dimensional stability diagrams showing the evolution of the threshold of the fold $F$ (black squares) and AH bifurcation points (a) $H_2-H_3$ and (b) $H_4-H_7$ in $(g,\,\alpha)$ plane for (a) $\Delta \varphi=0$ (red triangles) and (b) $\Delta \varphi=\pi$ (blue triangles). A BS is stable (a) in the area bounded by the lines $F$, $H_2, H_3$ and (b) between the lines $H_4$, $H_5$ and below the lines $H_7$ and $H_6$.}
\label{fig:Ros1d_alpha} 
\end{figure}

In~\cite{GJ-PRA-17} the influence of both linewidth enhancement factors on the stability of a single LB was studied. In particular it was shown that the range of the stability of the LB increases toward higher $\alpha$ values and decreases for growing $\beta$. Figure~\ref{fig:Ros1d_alpha} represents the corresponding bifurcation diagram for the BS in the $(g,\alpha)$ plane for the fixed value of $\beta$ and $d$ for (a) $\Delta \varphi=0$ and (b) $\Delta \varphi=\pi$, respectively. One can see that in the first case the fold position $F$ (black squares) remains almost constant for increasing $\alpha$ and almost coincides with the AH $H_3$ line (red triangles). Hence, the stability region is bounded by the AH lines $H_2$ and $H_3$ and is getting wider for increasing $\alpha$ values. For the phase difference $\Delta \varphi=\pi$ the stability diagram is more complicated since more AH points influence the dynamics of the BS, see Fig.~\ref{fig:Ros1d_alpha}~(b). As in the case of $\Delta \varphi=0$, the fold position remains almost constant for increasing $\alpha$ and almost coincides with the line of the AH point $H_7$. Hence, for small values of $\alpha$, the BS is stable between the $H_7$ and $H_4$ lines. However for increasing $\alpha$ two other AH points $H_5$ and $H_6$ emerge. The corresponding lines are moving towards larger (smaller) gain values so that two stability regions emerge. However, the $H_6$ line collides with $H_7$ at some $\alpha$, so that beyond this point, only one region of stability between $H_5$ and $H_4$ lines exist.  

\begin{figure}[ht!]
\includegraphics[width=1.0\columnwidth]{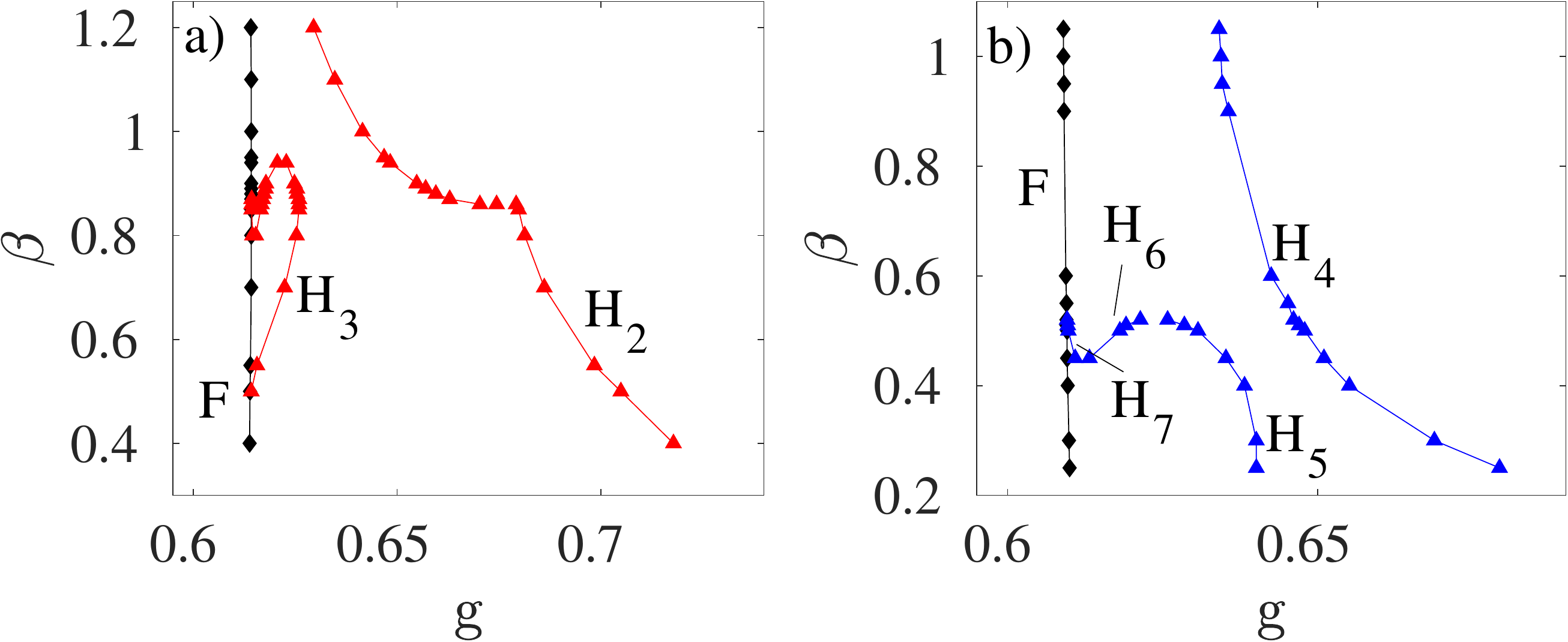}  
\caption{One-dimensional stability diagrams showing the evolution of the fold $F$ (black squares) and AH bifurcation points (a) $H_2-H_3$ and (b) $H_4-H_7$ in  $(g,\,\beta)$ plane for (a) $\Delta \varphi=0$ (red triangles) and (b) $\Delta \varphi=\pi$ (blue triangles). A BS is stable in the area bounded by the lines (a) $F$, $H_2, H_3$ and (b) $F$, $H_4$ and $H_5-H_7$.}
\label{fig:Ros1d_beta} 
\end{figure}

A similar stability diagram in $(g,\beta)$ plane for the fixed value of $\alpha$ and $d$ is shown in Fig.~\ref{fig:Ros1d_beta}. The width of the stability region is limited with either the fold $F$ or the corresponding AH lines ($H_3$ and $H_5-H_7$, respectively) on the left and the AH line $H_2$ ($H_4$) on the right. In both cases, the region of stability shrinks with increasing $\beta$.

\begin{figure}[ht!]
\includegraphics[width=1.1\columnwidth]{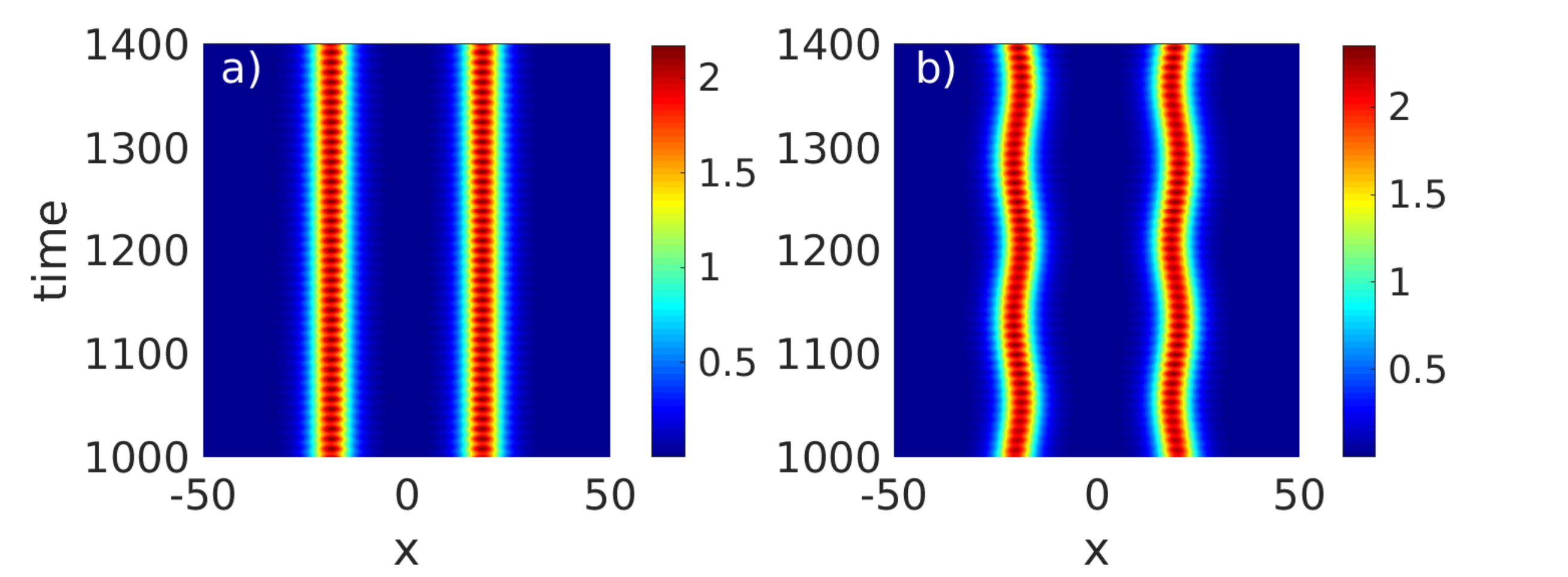}  
\caption{A cross-section of an oscillating BS of the two-dimensional Rosanov equation~\eqref{eq:Rosa1}-\eqref{eq:Rosa2} calculated on the two-dimensional domain $200\times 200$ with $512\times 512$ grid points for $\Delta \varphi$=$\pi$ at (a) $g=0.725$ and (b) $g=0.738$. The other parameters are as in Fig.~\ref{fig:RE1d_PD0}.}
\label{fig:Ros2D_1} 
\end{figure}

\paragraph*{Bound state dynamics in a 2D Rosanov equation}
In the two-dimensional case of Eqs.~\eqref{eq:Rosa1}-\eqref{eq:Rosa2} the dynamics of the BS solution is similar to the one-dimensional case: A radially-symmetric stationary BS solution exists over the wide range of the gain $g$ values and becomes unstable via several AH bifurcations, which morphology depends on the phase difference, the field diffusion and linewidth enhancement factors. Two examples of different oscillatory behavior calculated for $\Delta \varphi=\pi$ for two different gain values is shown in Fig.~\ref{fig:Ros2D_1}, where Fig.~\ref{fig:Ros2D_1} (a) represents the time evolution of a cross section of a BS of two LB that both oscillate in time keeping the distance between both LBs constant, whereas the panel (b) shows a BS with an additional oscillation in the distance. 


\paragraph*{Bound states of LB in the 2D Haus model}

\begin{figure}[ht!]
\includegraphics[width=1.0\columnwidth]{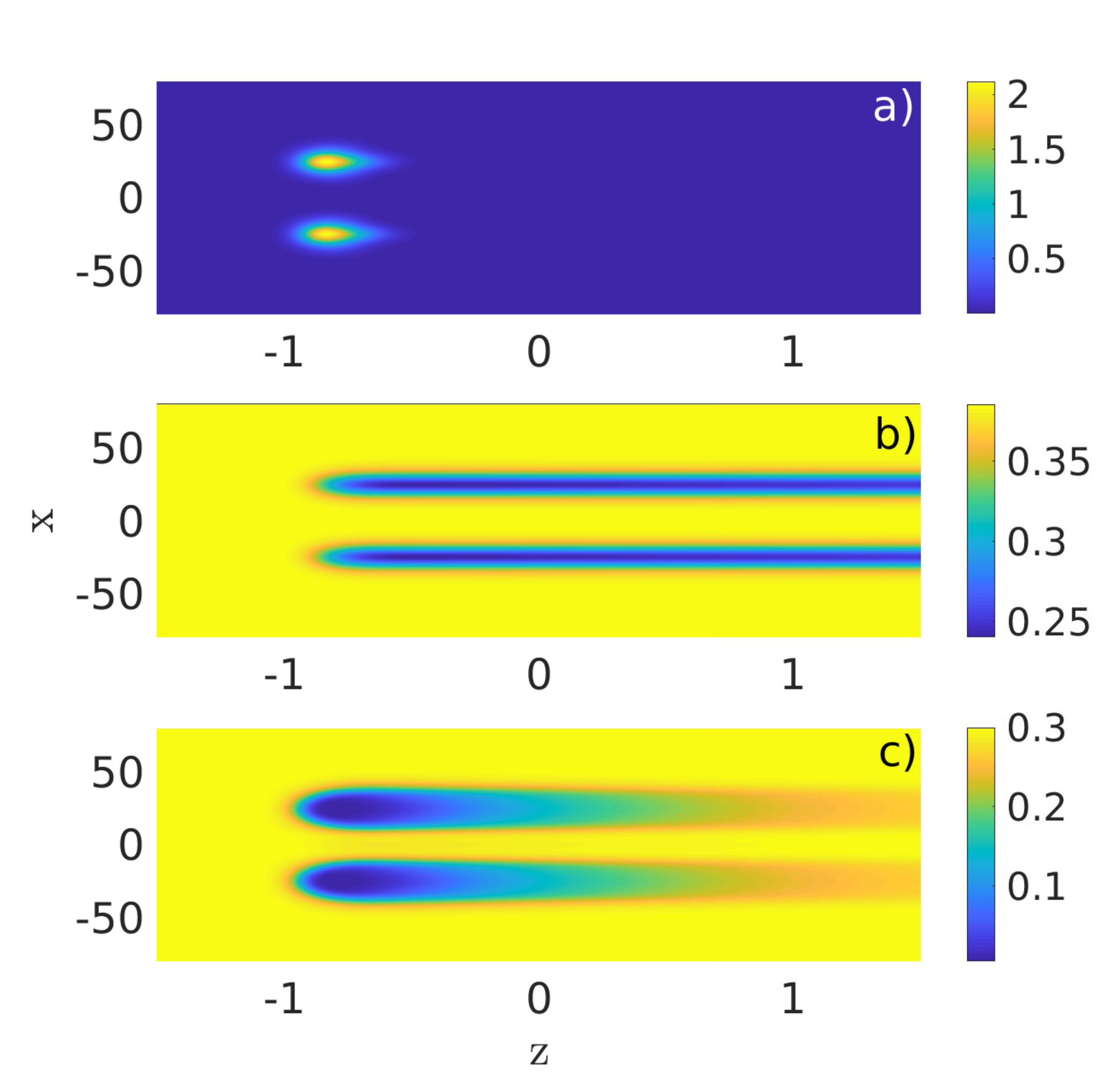}  
\caption{An exemplary numerical BS solution of Eqs.~\eqref{eq:VTJ1}-\eqref{eq:VTJ3} showing (a) the intensity profile $I$ (b) gain $G$ and (c) absorber $Q$ fields of the stable BS. The spatial component is in phase corresponding to $\Delta \varphi=0$. Other parameters are ($\gamma,\,\kappa,\,\alpha,\,\beta,\,G_0,\,\Gamma,\,Q_0,\,s,\,d,\,L_x,\,L_z,\,N_x, N_z$)=(40,\,0.8 \,1.5,\,0.5,\,0.736,\,0.04,\,0.3,\,30,\,0.03,\,300,\,3,\,512,\,512).}
\label{fig:HausSState} 
\end{figure}

In order to validate our approach, now we compare the results of the bifurcation analysis of the one-dimensional Rosanov model~\eqref{eq:Rosa1}-\eqref{eq:Rosa2} with the numerical predictions
of the two-dimensional Haus equations~\eqref{eq:VTJ1}-\eqref{eq:VTJ3}; The details of the numerical methods are given in~\cite{GJ-PRA-17}. We study Eqs. \eqref{eq:VTJ1}-\eqref{eq:VTJ3} in the regimes of LSs, that is, for values of the gain below threshold. In this regime, similar to the single LBs~\cite{}, BSs occur as multiple time scale objects as shown in Fig.~\ref{fig:HausSState}: Here, one notices that the optical component presented in Fig.~\ref{fig:HausSState}~(a) is much shorter than the material ones in Fig.~\ref{fig:HausSState}~(b),(c), where the gain $G=G(x,z)$ and absorber $Q=Q(x,z)$ fields are depicted.
One can see that the BS presented in Fig.~\ref{fig:HausSState} consists in two LBs, which longitudinal components correspond to two temporal LSs. In the transverse plane the pulses have weakly overlapping oscillatory tails, thus forming a BS with a certain distance $d$ and phase difference $\Delta \varphi$ between the pulses.  
\begin{figure}[ht!]
\includegraphics[width=1.1\columnwidth]{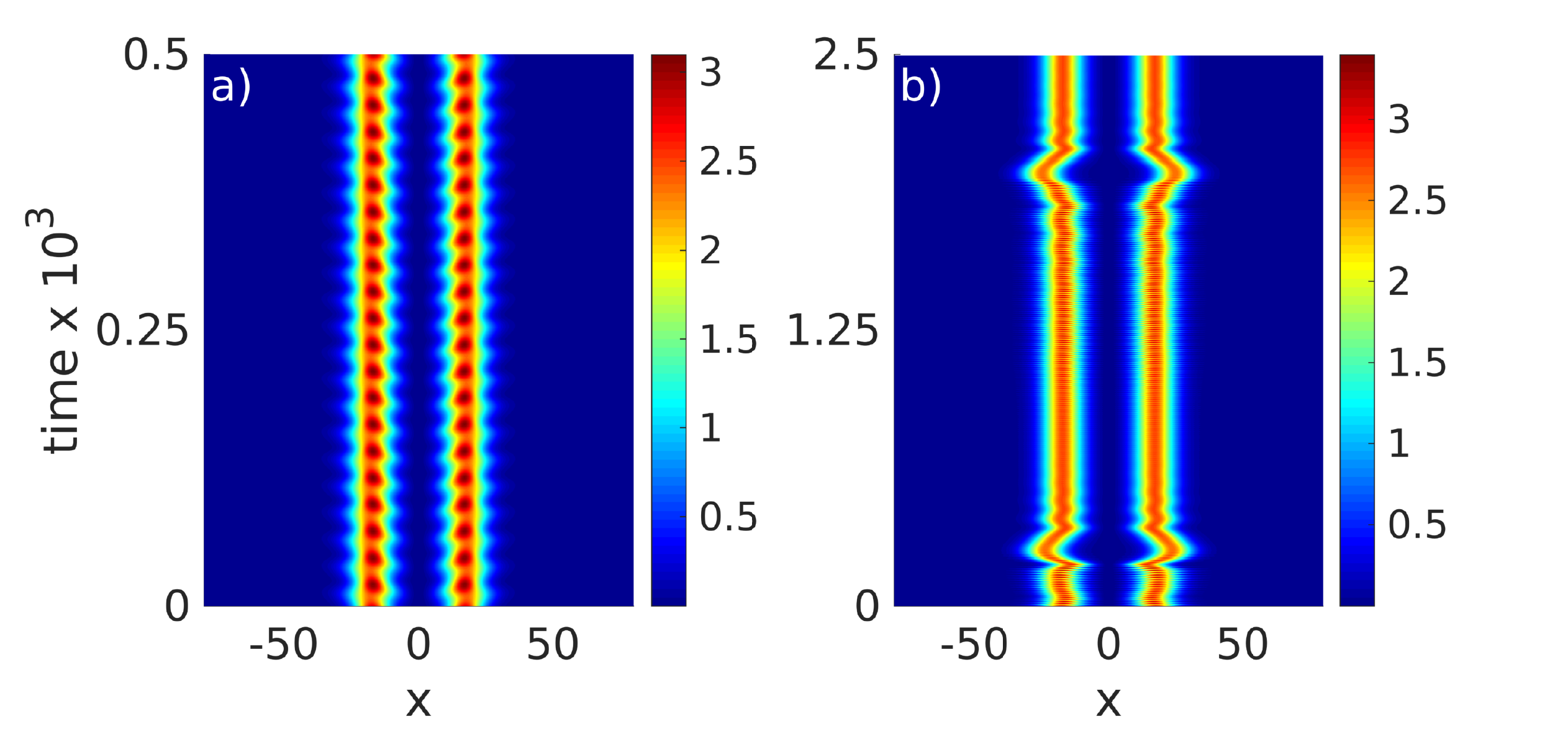}  
\caption{Spatio-temporal dynamics of a cross-section of a BS consisting of two LBs obtained from a direct numerical simulations of a two-dimensional Haus Eqs.~\eqref{eq:VTJ1}-\eqref{eq:VTJ3} for $\Delta \varphi=\pi$ and (a) $g=0.738$ and (b) $g=0.739$. While (a) demonstrates periodic intensity oscillations in time, b) contains additional oscillations in the distance between the bound LBs. Other parameters are the same as in Fig.~\ref{fig:HausSState}.}
\label{fig:HausOscilX} 
\end{figure}
To analyze the dynamical properties of these two-dimensional BS, a series of direct numerical simulations of Eqs.~\eqref{eq:VTJ1}-\eqref{eq:VTJ3} were conducted. They indicate that for both phase differences stable BSs become unstable via AH bifurcations leading to a complex oscillatory dynamics. The resulting dynamical regimes are similar to those obtained from the one-dimensional Rosanov Eqs.~\eqref{eq:Rosa1}-\eqref{eq:Rosa2}. One example is shown in Fig.~\ref{fig:HausOscilX}, where a temporal evolution of the BS's cross-section is shown for two different values of the gain and $\Delta \varphi=\pi$. One can see in Fig.~\ref{fig:HausOscilX}~(a) that the BS can, e.g., experience oscillation w.r.t. the center of each LBs, whereas the distance between them remains fixed in time (cf. Fig.~\ref{fig:REIMRos}~(d)) or can show complex quasi-periodic explosion-like oscillations as in Fig.~\ref{fig:HausOscilX}~(b) similar to the dynamics observed for the one-dimensional Rosanov model in Fig.~\ref{fig:ComplDyn}~(b). 

\section{Conclusion}

In conclusion, in this paper we discussed the dynamics and formation mechanisms of bound states consisting of light bullets. We have shown that the dynamics of a three-dimensional BS can be successfully approximated by a simplified model governing the dynamics of the transverse profile of the BS. The bifurcation analysis of this effective Rosanov equation allowed us to obtain guidelines regarding the existence and stability of the BS. Starting with the case of one spatial dimension, we have found that BSs corresponding to different phase differences can exist. While BSs corresponding to $\Delta \varphi=\{0,\,\pi\}$ can be stable in a range of system parameters, the stationary BSs with $\Delta \varphi=\dfrac{\pi}{2}$ are always unstable and moving BSs exist. This finding fits with the result found in~\cite{VKR_PRE01} where the dynamics of transverse autosolitons in bistable interferometers corresponding to the static saturated nonlinearity in Eqs.~\eqref{eq:Rosa1}-\eqref{eq:Rosa2} was studied. However, in our case for both $\Delta \varphi=\{0,\,\pi\}$ the existence of a threshold value of the field diffusion that bounds the stability region from below was demonstrated, whereas in~\cite{VKR_PRE01} the antiphase BS can be stable for vanishing values of the field diffusion. This threshold value also depends on the gain value and is in general higher then in $\Delta \varphi=0$ case. In addition, we have shown that as a function of the gain, the stability range of a BS is governed by the evolution of a fold or an AH bifurcation point, and an upper limiting point where the system develops an oscillating instability that can result in a complex oscillatory dynamics of the BS. These results are confirmed by direct numerical simulations of both two-dimensional Rosanov equation and two-dimensional Haus model. Note that in our analysis we were focused on the spatially instabilities of BSs of the LBs as governed by the transverse profile. However, more instabilities stemming from the temporal dynamics~\cite{SJG-PRA-18} are expected to influence the dynamics of BSs. The analysis and characterization of the full two-dimensional dynamics of the BSs of the Haus model can effectively be done employing the path-continuation methods. However, the multiscale nature of individual LBs and the additional transverse spatial dynamics makes the implementation technically involved and will be a topic of further studies. 

\section*{Acknowledgments}
S.G. thanks PRIME programme of the German Academic Exchange Service (DAAD) with funds from the German Federal Ministry of Education and Research (BMBF). J.J. acknowledge the financial support of the MINECO Project MOVELIGHT (PGC2018-099637-B-100 AEI/FEDER UE).


\end{document}